# Estimation of maximum possible trapped field in superconducting permanent magnets in 2D and 3D

**Víctor M. R. Zermeño[1], Algirdas Baskys[2], Shengnan Zou[1], Anup Patel[2] and Francesco Grilli[1k]**

[1] **Institute for Technical Physics, Karlsruhe Institute of Technology, Germany**

[2] **Applied Superconductivity and Cryoscience Group, Dept. Materials Science, Cambridge University, UK**

Email: victor.zermeno@kit.edu

**Abstract**

The ability of stacks of superconducting tapes to trap large magnetic fields makes them ideal candidates for creating powerful permanent magnets of compact size and mass. Experimentally, several techniques are used to trap the maximum possible field in a given practical application. However, regardless of the magnetization method used, there is a physical limit to the maximum magnetic field that a given superconducting magnet can trap. This limit is given by the geometric design, the particular superconducting material used and the temperature of operation. Knowing the maximum possible trapped field is important for device design as it provides an upper limit for applications such as magnetic bearings or rotating machinery. In this work we present a collection of finite element method (FEM) models in 2D and 3D capable of estimating the maximum trapped field of stacked tape superconducting magnets. The models are computationally fast and can be used to perform parametric studies with ease. For the case of square stacks tape magnets, various sizes are considered and their estimated maximum trapped field is compared with experimental results.

**1.- Introduction**

In recent years, stacks of superconducting tapes have achieved increased attention since they can be used as permanent magnets capable of trapping much larger fields than their non-superconducting counterparts. Given their mechanical and thermal stability, they offer advantages for compact and robust designs. Superconducting tapes with increasing critical current uniformity are currently being produced. Additionally, very detailed in-field characterization of superconducting tapes can be done [1]. This allows for reliable models of tape-stacked superconducting permanent magnets and their applications.

In this context, estimating the maximum field a superconducting magnet can trap is of high importance for design and optimization. Several works have addressed the problem of modelling the more complicated task of simulating the entire magnetization process of superconducting bulks. A recent comprehensive review [2] addressed the subject of modelling magnetization of bulk superconductors. There, several methods ranging from analytic approximations to 3D time dependent FEM models are described. The interested reader is encouraged to look at the references therein for particular details as several of the modelling techniques described can also be used for superconducting stacks.

Although tape-stacked superconducting permanent magnets are relatively new, a few works have already addressed their modelling and simulation. Some of said works aim to optimize the stack [3], to study the effect of the stabilizer [4], to optimize the magnetization process [5] [6] or even to simulate its performance in applications [7]. Most of the works previously mentioned intend to simulate the complete magnetization process. However, for many applications it is only necessary to know the maximum possible trapped field. One example of such modelling is given in [8], where a model considering a uniform current density is used to scale experimental data of a larger superconducting stack.

In this work we present a FEM model for estimating the maximum possible trapped field in a stack of tapes. One particular, but important feature of the model is that it can consider the dependence of the critical current density on the magnetic field even in the case of square or rectangular stacks. The model builds on previous 2D models [9] [10] [11] used to estimate the critical current of a superconducting device. In the present contribution, the model strategy used in [11] is extended to cover also the magnetization case. Analogously to estimating the critical current in the case of transport, here we consider the maximum possible trapped field in a superconducting bulk or stack. The cases of cylindrical, infinitely long and finite rectangular stacks are considered.

## 2.- Model description

Following the strategy introduced in [11], the present model assumes that all transients have mostly relaxed. Therefore, the current density $\boldsymbol{J}$ is assumed to have almost reached its steady state. This may seem a very strong assumption, as flux creep continuously demagnetizes the sample. However, said flux creep is expected to show a logarithmic decay rate. Therefore, although a proper steady state cannot be reached, shortly after the initial magnetization, flux creep leads to a very slow dynamic process. Several applications of magnetized stacks find use after this initial transitory period, and this explains the interest in estimating the maximum possible trapped field with a fast model that could be used for optimization purposes.

The present model considers the dependence on the local magnetic field $\boldsymbol{B}$ of the critical current density $J_c(\boldsymbol{B})$ and is built upon the following assumptions: current streamlines are known for the geometries considered and a homogenization approach is used to reduce the number of degrees of freedom and increase the computing speed.

Said homogenization approach is not a mere substitution of the superconducting stack with a homogenous isotropic bulk. Instead, it preserves the anisotropy of the stack by impeding any current flow from one superconducting layer to another. This is done by construction, by forcing the current to flow along prescribed streamlines.

In this work, the following geometries are considered: cylindrical stack of tapes, stack of infinitely long tapes, square and rectangular stacks of tapes. In the cases presented in this section, all stacks – centred at the coordinates' origin – are composed of 100 tapes whose surfaces are perpendicular to the z axis. The whole height of the stack is 5.5 mm.

For ease, and without loss of generality when comparing the four models, a tape with a critical current as the one reported in [12] was considered. If we denote the components of the magnetic flux density that are parallel and perpendicular to the tape's surface respectively by $B_\parallel$ and by $B_\perp$, $J_c(B_\parallel, B_\perp)$ can be expressed as follows:

$$J_c(B_\parallel, B_\perp) = \frac{J_{c0} f_{HTS}}{\left(1 + \sqrt{(k\, B_\parallel)^2 + B_\perp^2}/B_0\right)^\alpha}$$

where $J_{c0} = 47.5 \text{ GA/m}^2$, $k = 0.25$, $B_0 = 35 \text{ m}T$, and $\alpha = 0.6$. The remaining parameter, $f_{HTS} = 0.0182$, is the volume fraction of the superconducting material in the homogenized bulk. In what follows, we will only refer to this engineering current density. Actual values for the current density in the superconducting layers can be obtained by dividing the critical current density by $f_{HTS}$.

Cylindrical stack

The simplest case considered is a cylindrical stack of tapes (see **Figure 1**a). Here it is enough to assume that the superconducting layers carry a current density equal to $J_c(\boldsymbol{B})$ as shown in **Figure 1**b. This requires solving the non-linear problem of self consistently finding both the current density and the magnetic field. Using the magnetic vector potential $\boldsymbol{A}$ defined as $\boldsymbol{B} = \nabla \times \boldsymbol{A}$, this can be done by solving Ampere's law assuming a current source that is equal to $J_c(\boldsymbol{B})$:

$$\nabla \times \frac{1}{\mu} \nabla \times \boldsymbol{A} = J_c(\boldsymbol{B})\, \widehat{\boldsymbol{\varphi}}, \tag{1}$$

where $\widehat{\boldsymbol{\varphi}}$ is the unit vector pointing in the azimuthal direction. The simulated current density in the homogenized stack is shown in **Figure 1**c, and the corresponding magnetic flux density in **Figure 1**d. One can note that the point where the current density is highest (magnetic flux density is lowest) draws a circular kernel line.

Stack of infinitely long tapes

In a similar fashion, infinitely long stacks as the one shown in **Figure 2**a can be simulated using a 2D model that considers only a cross section of the long stack. As shown in **Figure 2**b, the magnetization condition is taken into account by considering that the superconducting material in the regions $y > 0$ and $y < 0$ has a critical current density of $J_c(\boldsymbol{B})$ and $-J_c(\boldsymbol{B})$ respectively. Again, using Ampere's law they can be modelled as:

$$\nabla \times \frac{1}{\mu} \nabla \times \boldsymbol{A} = sign(y) J_c(\boldsymbol{B}) \, \hat{\boldsymbol{\imath}}, \tag{2}$$

where $\hat{\boldsymbol{\imath}}$ is the unit vector in the x direction. The simulated current density in the homogenized stack is shown in **Figure 2**c, and the corresponding magnetic flux density in **Figure 2**d. One can now note that the kernel line denoting the point where the current density is highest (magnetic flux density is lowest) is much closer to the centre of the sample than in the case of a cylindrical stack.

Square and rectangular stacks of tapes

For the case of square and rectangular stacks as shown in **Figure 3**a, concentric rectangular paths for the current streamlines – denoted by $\Gamma_i$ in **Figure 3**b – are considered. This follows from the critical state model assumption for the magnetization of type-II superconductors with rectangular cross section given in [13]. If a constant critical current $J_c$ is used, said rectangular paths are enough to guarantee current conservation. However, the same does not hold when a non-constant critical current density $J_c(\boldsymbol{B})$ is considered. In this case, the local magnetic field along a given current streamline $\Gamma_i$ is not uniform and therefore $J_c(\boldsymbol{B})$ will take different values along a given streamline $\Gamma_i$. The point along $\Gamma_i$ where $J_c(\boldsymbol{B})$ is a minimum can be considered as a bottleneck for the current in that particular streamline. To guarantee current conservation, the minimum value of $J_c(\boldsymbol{B})$ along $\Gamma_i$ is used as the value of the current density. Again, this requires finding both the current density and the magnetic field that self-consistently solve Ampere's law. For a rectangular stack bounded by $|x| \leq a$ and $|y| \leq b$, the problem can be stated as:

$$\nabla \times \frac{1}{\mu} \nabla \times \boldsymbol{A} = \min_{\Gamma_i} \boldsymbol{J_c}(\boldsymbol{B}) \begin{Bmatrix} sign(x) \, \hat{\boldsymbol{\jmath}} & |y| < |x| + b - a \\ -sign(y) \, \hat{\boldsymbol{\imath}} & |y| > |x| + b - a \end{Bmatrix} \tag{3}$$

where $min_{\Gamma_i} \boldsymbol{J_c}(\boldsymbol{B})$ is the minimum value of $J_c(\boldsymbol{B})$ along each streamline $\Gamma_i$.

Figure 4a shows the computed magnetic flux density in the homogenized stack. The corresponding current density is shown in Figure 4b. The kernel line denoting the streamline of maximum current density follows a square path as prescribed. It is worth noting in Figure 4b that said kernel line is located close to the periphery of the sample, much like in the cylindrical stack mentioned before (see **Figure 1**c), and unlike in the case of an infinite long stack (see **Figure 2**c).

A comparison between the magnitudes of the critical current density function $J_c = |\boldsymbol{J_c}(\boldsymbol{B})|$ and the actual current density $J = |min_{\Gamma_i} \boldsymbol{J_c}(\boldsymbol{B})|$ in the central *x-y* plane of a square homogenized stack is shown in **Figure 5**. Here, it is important to note that $J_c$, depicts the maximum possible *local* value of $J$. However since the current is prescribed to follow concentric square paths, the minimum value of $J_c$ along each streamline will determine the actual current density $J$, effectively acting as a bottleneck.

A further insight into this bottleneck effect can be obtained from **Figure 6**. Here, a comparison of the current density inside one octant of three different rectangular stacks with aspect ratios of 1:1, 1:2 and 1:5 is shown. Although the overall profile of the current density is very similar for all rectangular samples, it is of interest to note that stacks with lower aspect ratios showed higher local the current density values. This can be better understood by considering the limit cases of an infinitely long stack (see **Figure 2**c) and a square stack (see **Figure 4**b). In an infinitely long stack, no end-effects are considered; hence the current distribution does not experience any limitation due to the different $J_c$ profile – given the larger magnitude of the magnetic flux density – at the stack's ends. This end effect is most clearly seen in the square stack. These two effects mutually constrain the overall current trapped in the sample.

It is worth noting that although the model presented in this section is based on the FEM, other implementations by means of Biot-Savart's law can be implemented as in [14] for the case of estimating the critical current of conductors.

To test its efficacy when estimating the maximum trapped field of a square stack, the model was validated with experimental results. Stacks of various sizes where considered and the experimentally measured trapped field compared with the estimates given by the model.

### 3.- Experimental methods and techniques

Superconducting Tape

The High Temperature Superconductor (HTS) tape used for the trapped field measurements was produced by SuperPower Inc. conforming to specification SP12050 AP. The tape had a 12 mm width, 50 µm thick Hastelloy substrate and a 2 µm thick silver over-layer leading to overall thickness of ~55 µm. The manufacturer stated that the minimum critical current was 240 A, at 77 K and self-field over the length of the tape, however segments showed much higher Ic, as shown in **Figure 7**. The tape was cut into 12x12 mm segments to make stacks of various sizes for field-cooling and 40 mm long segments for goniometric critical current measurements.

Tape characterization

Goniometric Ic(B, θ) measurements were performed at 77 K in liquid nitrogen and a magnetic flux density range of 0 to 0.5 T. The critical current goniometer is described in [1] and experimental setup is detailed in [15]. The critical current was measured for magnetic field orientations θ between -180° and +180°, where θ = 0° corresponds to the field being parallel to the tape normal and θ = ±90° corresponds to the direction perpendicular to both the tape normal and the current transport direction.

Sample voltage was measured across 1 cm and the electric field criterion of 1 µV/cm was used to determine the critical current, shown in **Figure 7**a. The power law $E = E_0 (I/I_c)^n$ was fitted to the data to determine the critical current $I_c$ and the $n$-value.

Field Cooling of Stacks of Tape

Field cooling was performed at 77 K in liquid nitrogen for stacks with different number of layers of superconducting tape. Orientations of each layer were kept identical through the stack. Magnetic field was applied by a Walker Scientific HV-4H electromagnet and ramped down from 1 T at a rate of 0.1 T/s to ensure that the sample is fully magnetized. The trapped field was recorded 2 minutes after the end of field cooling to ensure that the flux creep does not affect the measurement results significantly. The Hall probe was placed 1 mm above the surface of the stack, centred in the middle. For a more detailed description of the experimental setup, see [15].

**4.- Experimental results and model verification**

The tape was characterized using a 1 µV/cm criterion for the critical electric field. However, this value cannot be related to persistent magnetization currents as in the case of the magnetized stacks here considered. At such high electric field, substantial dissipation already occurs in the superconductor. Therefore, the experimentally measured values of $I_c$ at 1 µV/cm should be scaled down to better represent a state where most dissipation has already taken place. This rescaling can be achieved by considering lower criteria for the critical electric field. However, this is experimentally challenging and not-practical as it would require performing the characterization in an extremely low noise environment. Figure 8 shows a current – voltage characteristic modelled using a power law (black line) together with the expected signal once a noise level of 0.1 µV/cm is considered (blue line). It is clear that lower criteria for the critical electric cannot be easily obtained as they would be concealed under the noise level. For this reason, we have scaled the critical current using a power law with exponent n=14 to a lower electric field criterion. This rescaling of the critical current uses the critical electric field as a fitting parameter and can be easily justified considering the following two assumptions: First, that flux creep is expected to dominate the dynamics of the system for larger critical electric field criteria and second, that once the initial dissipation has taken place, the flux creep is expected to become a very slow dynamic process.

As shown in Figure 9, using a criterion of 3.2 nV/cm gave excellent agreement with experimental data for a large number of stacks considered. The scaled down values of $I_c$ using this criterion are shown in Figure 7b. The $J_c(\boldsymbol{B})$ obtained with this new criterion was used as a parameter to match the measurements performed 2 minutes after the end of the field cooling. For reference, estimated values for the trapped field using other electric field criteria are also shown in Figure 9.

**5.- Conclusion**

In this work, we presented FEM models for the computation of the maximum possible trapped field in stacked-tape superconducting permanent magnets of various shapes. The models are based upon the assumption that the current streamlines follow known paths; either circles in cylindrical stacks, straight lines in stacks of infinitely long tapes, or a series of concentric rectangles for the case of rectangular (and square) stacks.

The models were used to estimate the maximum possible trapped field in a square stack that was magnetized following the field cooling method. Excellent agreement with experimental data was obtained once the in-field critical current of the tapes was scaled down to fit the measured values of trapped field.

Beyond its direct applicability to simulate a given configuration and to provide further understanding of the properties of superconducting permanent magnets, the proposed stationary model could be used for optimization purposes as several designs can be rapidly simulated and their performance evaluated.

**Aknowledgements**

This work was in part supported by the Engineering and Physical Sciences Research Council, U.K.

**References**

[1]     S. C. Hopkins, M. Woźniak, B. A. Glowacki, Y. Chen, I. Kesgin, and V. Selvamanickam, "Two-axis Magnetic Field Orientation Dependence of Critical Current in Full-Width REBCO Coated Conductors," *Phys. Procedia*, vol. 36, pp. 582–587, 2012.

[2]     M. D. Ainslie and H. Fujishiro, "Modelling of bulk superconductor magnetization," *Supercond. Sci. Technol.*, vol. 28, no. 5, p. 053002, May 2015.

[3]     A. Patel and B. a Glowacki, "Optimisation of composite superconducting bulks made from (RE)BCO coated conductor stacks using pulsed field magnetization modelling," *J. Phys. Conf. Ser.*, vol. 507, no. 2, p. 022024, May 2014.

[4]     A. G. Page, A. Patel, A. Baskys, S. C. Hopkins, V. Kalitka, A. Molodyk, and B. A. Glowacki, "The effect of stabilizer on the trapped field of stacks of superconducting tape magnetized by a pulsed field," *Supercond. Sci. Technol.*, vol. 28, no. 8, p. 085009, 2015.

[5]     S. Zou, V. Zermeno, and F. Grilli, "Simulation of Stacks of High Temperature Superconducting Coated Conductors Magnetized by Pulsed Field Magnetization Using Controlled Magnetic Density Distribution Coils," *IEEE Trans. Appl. Supercond.*, pp. 1–1,

Sep. 2016.

[6]  S. Zou, V. M. R. Zermeño, A. Baskys, A. Patel, F. Grilli, and B. A. Glowacki, "Simulation and experiments of Stacks of High Temperature Superconducting Coated Conductors Magnetized by Pulsed Field Magnetization with Multi-Pulse Technique," *Supercond. Sci. Technol.*, vol. Submitted, 2016.

[7]  F. Sass, G. G. Sotelo, R. de A. Junior, and F. Sirois, "H-formulation for simulating levitation forces acting on HTS bulks and stacks of 2G coated conductors," *Supercond. Sci. Technol.*, vol. 28, no. 12, p. 125012, 2015.

[8]  A. Patel, K. Filar, V. I. Nizhankovskii, S. C. Hopkins, and B. a. Glowacki, "Trapped fields greater than 7 T in a 12 mm square stack of commercial high-temperature superconducting tape," *Appl. Phys. Lett.*, vol. 102, no. 10, p. 102601, 2013.

[9]  M. Majoros, B. A. Glowacki, and A. M. Campbell, "Critical current anisotropy in Ag/(Pb,Bi) 2 Sr 2 Ca 2 Cu 3 O 10+ x multifilamentary tapes: influence of self-magnetic field," *Supercond. Sci. Technol.*, vol. 14, no. 6, pp. 353–362, Jun. 2001.

[10] F. Gömöry and B. Klinčok, "Self-field critical current of a conductor with an elliptical cross-section," *Supercond. Sci. Technol.*, vol. 19, no. 8, pp. 732–737, Aug. 2006.

[11] V. Zermeño, F. Sirois, M. Takayasu, M. Vojenciak, A. Kario, and F. Grilli, "A self-consistent model for estimating the critical current of superconducting devices," *Supercond. Sci. Technol.*, vol. 28, no. 8, p. 085004, Aug. 2015.

[12] F. Grilli, F. Sirois, V. M. R. Zermeno, and M. Vojenciak, "Self-Consistent Modeling of the Ic of HTS Devices: How Accurate do Models Really Need to Be?," *IEEE Trans. Appl. Supercond.*, vol. 24, no. 6, pp. 1–8, Dec. 2014.

[13] E. H. Brandt, "Electric field in superconductors with rectangular cross section," *Phys. Rev. B*, vol. 52, no. 21, pp. 15442–15457, Dec. 1995.

[14] V. Zermeno, F. Grilli, and S. Quaiyum, "Open Source Codes for Computing the Critical Current of Superconducting Devices," *IEEE Trans. Appl. Supercond.*, vol. PP, no. 99, pp. 1–1, 2016.

[15] A. Baskys, A. Patel, S. Hopkins, and B. Glowacki, "Modeling of trapped fields by stacked (RE)BCO tape using angular transversal field dependency," *IEEE Trans. Appl. Supercond.*, vol. 26, no. 3, pp. 1–1, 2016.

**Figures**

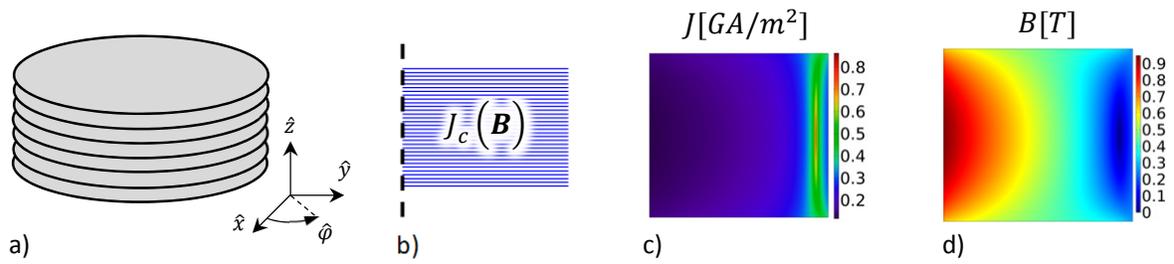

Figure 1. a) Cylindrical stack of tapes. b) The axisymmetric model assumes the whole superconducting region (denoted by blue lines) to have a critical current density equal to $\mathbf{J_c(B)}$. Profiles corresponding to the current density (c) and magnetic flux density (d) in a homogenized stack of 100 tapes.

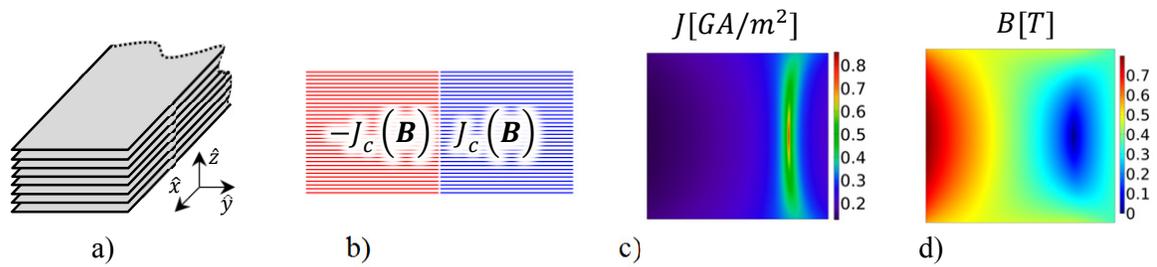

Figure 2. a) Stack of infinitely long tapes. b) The 2D model assumes the whole superconducting region to have a critical current density equal to $-\mathbf{J_c(B)}$ (red lines) and to $\mathbf{J_c(B)}$ (blue lines). Profiles corresponding to the current density (c) and magnetic flux density (d) in a homogenized stack of 100 tapes.

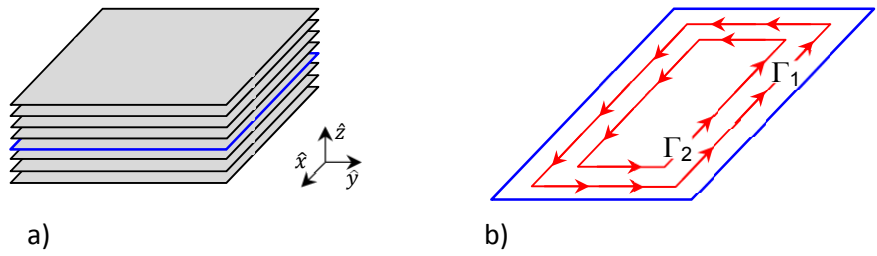

a) b)

Figure 3. a) Rectangular stack of tapes. b) Current stream lines are assumed to be concentric rectangles denoted by the closed paths $\Gamma_i$.

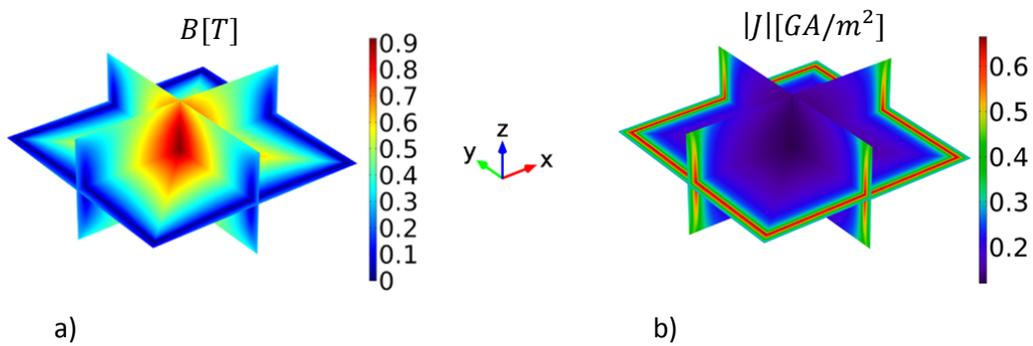

a) b)

Figure 4. Magnetic flux density (a) and current density (b) in a homogenized 5.5 mm-high, 12 mm-wide square stack composed of 100 tapes. The plots are composed of 3 slice planes intersecting the center of the stack.

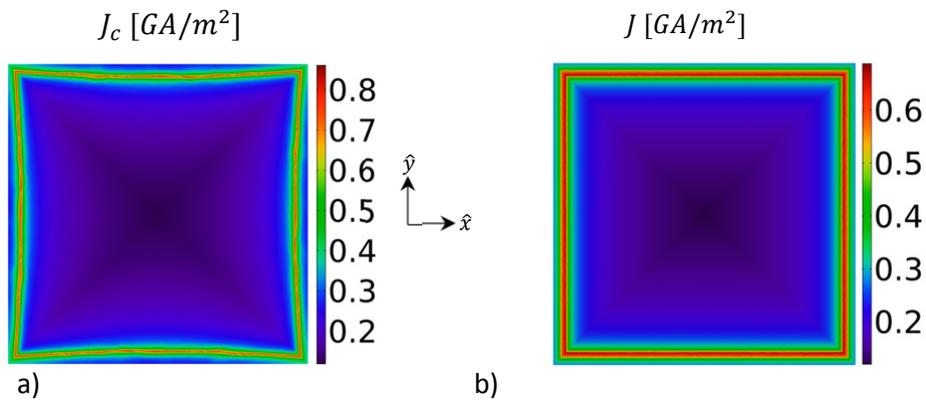

a) b)

Figure 5. Comparison between a) $\mathbf{J_c}$ and b) $\mathbf{J}$ at the center of a homogenized 5.5 mm-high, 12 mm-wide square stack composed of 100 tapes.

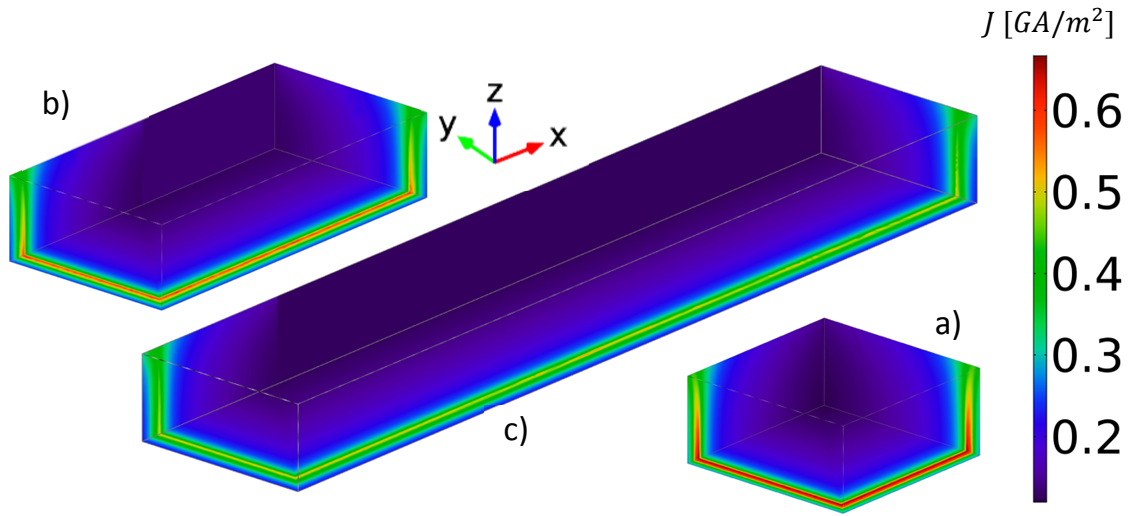

Figure 6. Effect of aspect ratio. Cross-section plot of current density in three 5.5 mm-high, 12 mm-wide rectangular homogenized stacks composed of 100 tapes of different aspect ratios: a) 1:1 b) 1:2 c) 1:5. Only one octant is plotted.

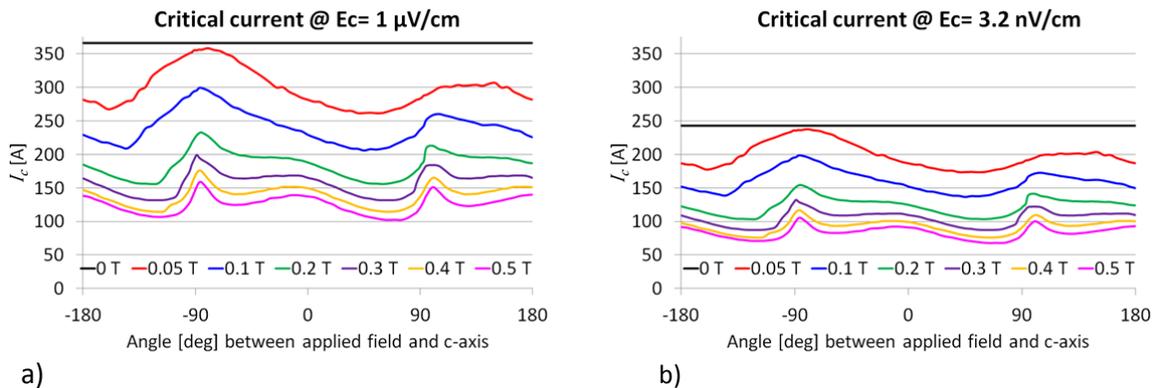

Figure 7. Goniometric Critical Current Measurements. a) Measured critical current values using the 1 µV/cm critical electric field criterion. b) Scaled critical current values using the 3.2 nV/cm critical electric field criterion. The scaling allows considering a state where the dissipation has been considerably reduced.

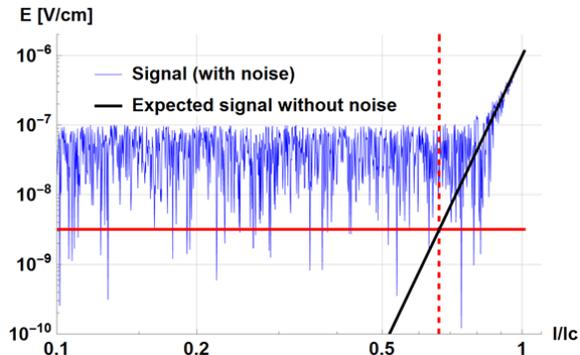

Figure 8. Current – voltage characteristic modelled via a power law with exponent n = 14. Expected measured signal for a noise level of 0.1 µV/cm (blue line). Expected signal without noise (black line). The red gridlines denote the 3.2 nV/cm criterion (solid line) and the corresponding critical current (dashed line).

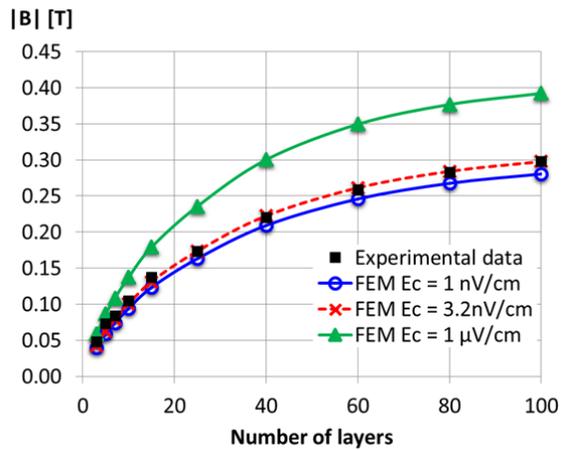

Figure 9. Experimentally measured trapped field 1 mm above the stack's surface. FEM simulations for different critical electric field criteria Ec are also shown for reference. The best agreement with experimental values was obtained with Ec = 3.2 nV/cm.